\documentclass[preprint,showpacs,showkeys,nofootinbib,preprintnumbers,floatfix]{revtex4}
\usepackage{amsmath,epsfig}

\begin{document}

\preprint{SNUTP 03-017}

\title{\Large\bf Exit from inflation and a paradigm for vanishing cosmological
constant in self-tuning models}

\vskip 3cm 

\author{Jihn E. Kim$^{(a,b)}$\footnote{jekim@phyp.snu.ac.kr} and
Hyun Min Lee$^{(b)}$\footnote{minlee@th.physik.uni-bonn.de}}
\address{$^{(a)}$School of Physics and Center for
Theoretical Physics, Seoul National University, Seoul
151-747, Korea\\
$^{(b)}$Physikalisches Institut,
Universit\"at Bonn, Nussallee 12, D53115, Bonn, Germany}

\begin{abstract}
We propose a paradigm for the inflation and the vanishing cosmological
constant in a unified way with the self-tuning solutions
of the cosmological constant problem. Here, we consider  
a time-varying cosmological constant 
in self-tuning models of the cosmological constant. 
As a specific example, we demonstrate it with a 3-form field in 5D.
\end{abstract}

\keywords{cosmological constant,self-tuning,inflation,brane}

\pacs{98.80.Es,98.80.C,12.25.Mj}

\maketitle

\section{Introduction}

Ever since the inflationary idea has been proposed, how
the universe settles after the inflationary period at the vacuum 
with the vanishing vacuum energy has been a dream to be solved 
but postponed until the solution of the cosmological constant
problem is known. In 4 dimensional(4D) field theory models, it is
known that there is no solution for the cosmological constant
problem~\cite{review}. One must go beyond 4D 
to find a clue to the solution of the cosmological
constant problem. In this sense, the Randall-Sundrum(RS) type models,
in particular the RS-II type models~\cite{rs2} are of great interest toward
a clue toward a vanishing cosmological constant.

Indeed, a few years ago solutions of the cosmological constant 
problem have been tried under the name of self-tuning 
solutions~\cite{kachru,nilles,kkl1,kl}. In the late 1970's and early
1980's it was called the solutions with an undetermined integration 
constant(s). The old self-tuning solutions looked for flat space
solutions whether or not it accompanies the de Sitter(dS) space and/or
anti-de Sitter(AdS) space solutions. This kind of old
self-tuning solutions is called {\it weak self-tuning solutions.}
On the other hand, recently it has been tried to find a self-tuning
solution without allowing the nearby dS and AdS space solutions~\cite{kachru}.
This kind of new self-tuning solutions can be called {\it strong
self-tuning solutions.} However, there seems no example for the 
strong self-tuning solution~\cite{nilles}.

The first example for the self-tuning solution is obtained in 5D with the
three index antisymmetric tensor field $A_{MNP}$, with the
$1/H^2$ in the action where $H_{MNPQ}=\epsilon^{MNPQ}\partial_MA_{NPQ}$
~\cite{kkl1,kkl2}. Certainly, this action has a few unsatisfactory 
features, but it renders an example for the existence of weak
self-tuning solutions, and provides possible physics behind the
self-tuning solutions. One such example is the existence of the 
region of parameter space where only the dS solutions are allowed,
which can be used for the period of inflation~\cite{kim}. 
From this example, we can envision a unification of the ideas 
of inflation in the early universe, presumably at the GUT era, 
and the solution of the cosmological constant problem.\footnote{The
present tiny vacuum energy of order $(0.003\ {\rm eV})^4$ is expected
to be understood by another independent mechanism such as by the
existence of quintessence.} Weak self-tuning solutions
have been tried with a string-inspired Gauss-Bonnet action with some
fine-tuning between bulk and/or brane parameters~\cite{binetruy}, 
and in models with brane gravity~\cite{kyae}. 
In view of the existence of a few weak
self-tuning solutions, therefore, the time is ripe to consider 
a unified view on the inflation and vanishing cosmological constant 
now even though there has not appeared yet a universally accepted
self-tuning solution.

In this vein, we try to find out time-dependent solutions of the
cosmological constant for a simplified step function change(with 
respect to $t$) of the
cosmological constant. We have tried this kind of time-dependent
step function for the brane tension before to show the existence of 
a flat space to another flat space solution in case the spontaneous
symmetry breaking changes the vacuum energy of the
observable sector~\cite{kkl2}. Our motivation in this paper 
is to see the time-dependent curvature change. However, 
the closed form dS solutions are difficult to find out. In fact, 
there has not appeared any closed form dS solution connected to a 
self-tuning solution. A time-dependent curvature solution is
even more difficult to obtain. Therefore, in this paper we just try 
to show the existence of such time dependent solutions of the curvature
and put forward a paradigm how the
cosmological constant can become zero after the inflationary era.
If there appears a universally accepted weak self-tuning solution in
the future, the solution of the cosmological constant problem can be
realized in this way.

There exist two examples of the closed form weak self-tuning 
solutions~\cite{kkl1,kl}. In this paper, we try to show the paradigm
with the weak self-tuning solution obtained with the 
$1/H^2$ term by Kim, Kyae, and Lee(KKL)~\cite{kkl1,kkl2}.

The unified view of the inflation and vanishing cosmological
constant is realized in the following way. The universe starts
with the parameters which allow only the dS space solutions~\cite{kim}.
Let us call this dS-only region {\it the D-region}.
In this phase there results a sufficient inflation. The inflationary
potential tried in 4D field theory models is the 4D potential
at the brane located at $y=0$ in the RS-II models. The observable
sector fields are localized at the $y=0$ brane. When the brane
tension becomes sufficiently small, but not necessarily zero,
the parameters enter into the region where the flat space,
de Sitter space and anti-de Sitter 
space solutions are allowed~\cite{kkl2}.
Let us call this flat space allowing region {\it the F-region}.
Then, we may consider an initial condition after exiting
from the {\it D-region} is a dS space solution in the
{\it F-region}. Since the flat space, dS space and AdS space solutions
are allowed in the F-region, we look for a time-dependent solution in
the F-region. In particular, we look for the curvature changing
solutions. If the effective 4D curvature tends to zero as 
$t\rightarrow\infty$,

\def\Cc{$\bar\Lambda_{\rm eff}$}
\begin{equation}\label{tdep} 
\bar\Lambda_{\rm eff}\propto \frac{1}{t^p}
\end{equation}
where $p>2$, then we obtain a reasonable solution for the cosmological
constant problem. It is necessary to require $p>2$ so that the
radiation dominated phase of the standard Big Bang cosmology commences
after inflation. The solution of the cosmological constant
problem can be realized with (\ref{tdep}) in weak self-tuning models.
If we cannot determine (\ref{tdep}) classically
as a solution of equations of motion, a quantum mechanical probabilistic
determination can be given. In this sense, Baum and Hawking's
probabilistic interpretation~\cite{hawking} is clearly envisioned 
in this 5D example. 

If we find a strong self-tuning solution, one must also
show the existence of the D-region to accomodate the inflationary era. 
Otherwise, it is not cosmologically successful. 

\section{Time-dependent curvature in weak self-tuning model}

As a prototype example of the weak self-tuning model, we consider
the KKL model~\cite{kkl1,kkl2}. Here, a three-form field 
$A_{MNP}$($M,N,P=0,1,\cdots,4$)\cite{kkl1,kkl2} is introduced.
In this model, it has been shown~\cite{kim} that
there exists a band of brane tension $\Lambda_1$, allowing only the
dS solutions, $|\Lambda_1|>\sqrt{-6\Lambda_b}(\equiv D-region)$, 
where $\Lambda_b$ is the bulk cosmological constant. 
On the other hand, both 4D flat 
and maximally curved($dS_4$ and $AdS_4$ spaces) solutions are allowed for 
$|\Lambda_1|<\sqrt{-6\Lambda_b}(\equiv F-region)$~\cite{kkl1,kkl2}.  
Thus, the KKL model has the ingredient needed for
inflation in the weak self-tuning model.\footnote{ 
Other self-tuning solutions with different forms for the action of 
$H_{MNPQ}=\partial_{[M}A_{NPQ]}$ have been also considered~\cite{kl}.}

In the KKL model, let us proceed to show a time-varying 4D cosmological 
constant.\footnote{This work has been reported at a recent 
conference~\cite{cosmo03}} Because of the difficulty of obtaining a closed form for
the $t$-dependent solution, we consider just the
the instantaneous transition between two different 
$dS_4$ curvature scales. 

The 5D action considered in the KKL model is
\begin{eqnarray}
S=\int d^4 x \int dy \sqrt{-g}\bigg(\frac{1}{2}R-\Lambda_b
+\frac{2\cdot 4!}{H^2}-\frac{\sqrt{-g_4}}{\sqrt{-g}}\Lambda_1\delta(y)\bigg)
\end{eqnarray}
where $g$, $g_4$ are 5D and 4D metric determinants, $H^2=H_{MNPQ}H^{MNPQ}$, and 
$\Lambda_b$, $\Lambda_1$ are bulk and brane cosmological constants, 
respectively. Henceforth, we use the dimensionless unit for
the fundamental scale, $M=1$. The fundamental unit $M$ can be
reintroduced when needed. 
Then, the ansatz for the $dS_4$ solution is  
\begin{eqnarray}
ds^2&=&\beta^2(y)(-dt^2+e^{2\sqrt{\bar\Lambda}t}\delta_{ij}dx^i dx^j)+b^2dy^2,
\\
H_{\mu\nu\rho\sigma}&=&\sqrt{-g}\epsilon_{\mu\nu\rho\sigma}f(y), \ \ \
H_{5ijk}=0
\end{eqnarray}
where $\bar\Lambda$ is the $dS_4$ curvature, and $b$ is a constant,
and $f^2(y)=2A/\beta^8(y)$ with an integration constant $A$. For 
this ansatz, the 4D curvature $\bar\Lambda$ is constant.
The (55) Einstein equation in the KKL model gives 
the governing equation of $\beta$\cite{kkl2,kim} 
\begin{eqnarray}
\frac{1}{b}\beta'=\pm\sqrt{{\bar k}^2+k^2\beta^2-Q^2\beta^{10}}\label{gv}
\end{eqnarray}  
where 
\begin{eqnarray}
k=\sqrt{-\frac{\Lambda_b}{6}}, \ \ {\bar k}^2={\bar\Lambda}, 
\ \ Q=\sqrt{\frac{1}{6A}}. 
\end{eqnarray}
Moreover, the boundary condition for the warp factor at $y=0$ is given by 
\begin{eqnarray}
\frac{\beta'}{\beta}\bigg|_{y=0^+}=-\frac{b}{6}\Lambda_1,\label{BC}
\end{eqnarray}
Then, if we take the negative sign on the RHS of Eq.~(\ref{gv}) 
for a positive $\Lambda_1$, the warp factor becomes 
$\beta(y)=\beta(b(-|y|+c);{\bar k}^2)$ with a positive
integration constant $c$. 
Even if the exact form for $\beta$ was not obtained, it has been shown 
numerically that
there always exists a $dS_4$ solution in the F-region~\cite{kkl2,kim}. 
Note that there is the repetition of bulk horizons 
among which only the first horizon 
at $y=c$ is causally connected to the observer at $y=0$ and the length 
scale $b c$ can be considered as the size of extra dimension. 
  
Let us consider the instantaneous change of $b(t)$ at $t=t_0$ 
as\footnote{This form was also considered for maintaining 
the flat solution with a changing brane tension in Ref.~\cite{kkl2}.}
\begin{eqnarray}
b(t)=(b_f-b_i)\theta(t-t_0)+b_i.
\end{eqnarray}
Then, the warp factor has a form of 
$\beta[|y|,t]=\beta[b(t)(-|y|+c);{\bar k}^2(t)]$
with time-dependent ${\bar k}(t)$. 
For a constant brane tension $\Lambda_1$,
the boundary condition (\ref{BC}) at $y=0$ reads
the time dependence of ${\bar k}^2(t)$ as
\begin{eqnarray}
{\bar k}^2(t)\beta^{-2}(0,t)+k^2 -Q^2\beta^8(0,t)=k_1^2 \label{bc}
\end{eqnarray}
where $k_1=\Lambda_1/6$.
Thus, ${\bar k}^2(t)$ is a function of $\theta(t-t_0)$ to be determined from 
knowing the exact form of $\beta$. Anyway, ${\bar\Lambda}(t)$ has the 
initial value ${\bar\Lambda}_i$ in terms of $b_i c$ 
and the final value ${\bar\Lambda}_f$ in terms of $b_f c$ via Eq.~(\ref{bc}). 

In fact, the brane value of $\beta$ and ${\bar k}^2$ consistent
with the boundary condition (\ref{bc}) are 
\begin{eqnarray}
\beta^2_i(0)>\beta^2_f(0), \ \ \ {\bar k}^2_i>{\bar k}^2_f,\label{ineq1}  
\end{eqnarray}
or
\begin{eqnarray}
\beta^2_i(0)<\beta^2_f(0), \ \ \ {\bar k}^2_i<{\bar k}^2_f,\label{ineq2}
\end{eqnarray}
where $\beta_i(0)\equiv\beta(0,t<t_0)$, $\beta_f(0)\equiv\beta(0,t>t_0)$, 
and ${\bar k}^2_i\equiv {\bar k}^2(t<t_0)$, 
${\bar k}^2_f\equiv {\bar k}^2(t>t_0)$. 
On the other hand, by integrating Eq.~(\ref{gv}) from $y=0$ 
to the first bulk horizon $y_h$ where $\beta=0$,
we get the bulk horizon size as
\begin{eqnarray}
b(t)c=\int^{\beta(0,t)}_0\frac{dx}{\sqrt{{\bar k}^2(t)
+k^2 x^2-Q^2 x^{10}}}.\label{horizon}
\end{eqnarray}
Inserting ${\bar k}^2(t)$ of Eq.~(\ref{bc}) into Eq.~(\ref{horizon}) 
in terms of $\beta(0,t)$
and making a change of integral variable with $x'=x/\beta(0,t)$, 
we can rewrite Eq.~(\ref{horizon}) as
\begin{eqnarray}
b(t)c=\int^1_0 \frac{dx'}{\sqrt{k_1^2-k^2(1-x^{\prime 2})
+\beta^8(0,t)Q^2(1-x^{\prime 10})}}. \label{horizon1}
\end{eqnarray}
Therefore, with the inequalities of Eqs.~(\ref{ineq1}) and (\ref{ineq2}), 
we find that the change of $b$ is $b_i<b_f$ for ${\bar k}^2_i>{\bar k}^2_f$
and $b_i>b_f$ for ${\bar k}^2_i<{\bar k}^2_f$. In other words, 
a larger(smaller) 4D cosmological constant gives a smaller(larger) 
bulk horizon size. 
 
From the ansatz for components of $H$ as
\begin{eqnarray}
H^{\mu\nu\rho\sigma}=\frac{1}{\sqrt{-g}}\epsilon^{\mu\nu\rho\sigma 5}
\partial_5\sigma, \ \ \
H^{5ijk}=\frac{1}{\sqrt{-g}}\epsilon^{5ijk0}\partial_0\sigma,
\end{eqnarray} 
the solution for $\sigma$ with time-dependent $b(t)$ is given from the static
$b$ case as
\begin{eqnarray} 
\sigma(y,t)=b^2(t)\sqrt{2A}\int dy \beta^{-4}(b(t)(-|y|+c);{\bar k}^2(t)).
\end{eqnarray}
Therefore, we get the time derivative of $\sigma$ as
\begin{eqnarray}
{\dot\sigma}=\sqrt{2A}\bigg[b{\dot b}(2\int dy \beta^{-4}+\beta^{-4})
+b^2{\dot{\bar k}}\int dy \frac{\partial \beta^{-4}}{\partial{\bar k}}\bigg]
\end{eqnarray} 
where there appear $\delta(t-t_0)$ terms due to ${\dot b}$ and $\dot{\bar k}$. 
Then, we find that the field equation for $H$ is satisfied with this 
time-dependent $\sigma$ even at $t=t_0$ as has been done 
in the flat case\cite{kkl2}. 
  
Moreover, the time derivative terms of $\beta$,$b$ and ${\bar k}$ 
in $a\equiv e^{{\bar k}t}$ in the Einstein 
equations are cancelled by the bulk matter fluctuation around the vacuum,
$T^{(m)M}\,_N\equiv {\rm diag}(-\rho,p,p,p,p_5)$, as
\begin{eqnarray}
\rho&=&\frac{3}{\beta^2}\bigg(2\frac{\dot a}{a}\frac{\dot\beta}{\beta}
+\frac{\dot a}{a}\frac{\dot b}{b}+\bigg(\frac{\dot\beta}{\beta}\bigg)^2
+\frac{\dot\beta}{\beta}\frac{\dot b}{b}
+2{\bar k}{\dot{\bar k}}t+{\dot{\bar k}}^2 t^2\bigg),\\
p&=&-\frac{1}{\beta^2}\bigg(4\frac{\dot a}{a}\frac{\dot\beta}{\beta}
+2\frac{\ddot\beta}{\beta}+\frac{\ddot b}{b}
-\bigg(\frac{\dot\beta}{\beta}\bigg)^2
+\frac{\dot\beta}{\beta}\frac{\dot b}{b}
+2\frac{\dot a}{a}\frac{\dot b}{b}+4{\dot{\bar k}}+2{\ddot{\bar k}}t
+4{\bar k}{\dot{\bar k}}t+2{\dot{\bar k}}^2 t^2\bigg),\\
p_5&=&-\frac{3}{\beta^2}\bigg(3\frac{\dot a}{a}\frac{\dot\beta}{\beta}
+\frac{\ddot\beta}{\beta}+2{\dot{\bar k}}t+{\ddot{\bar k}}t
+4{\bar k}{\dot{\bar k}}t+2{\dot{\bar k}}^2 t^2\bigg).
\end{eqnarray}
Note that the bulk matter contributes only at $t=t_0$ with terms proportional
to $\delta(t-t_0)$, $\delta^2(t-t_0)$ and ${\dot\delta}(t-t_0)$. 
Since the condition $G_{05}=0$ is also satisfied, 
the 5D continuity equations for the bulk matter are automatically satisfied,
\begin{eqnarray}
{\dot\rho}+3\bigg(\frac{\dot a}{a}+\frac{\dot\beta}{\beta}\bigg)(\rho+p)
+\frac{\dot b}{b}(\rho+p_5)=0, \\
p'_5+3\frac{\beta'}{\beta}(p_5-p)+\frac{\beta'}{\beta}(\rho-p_5)=0.
\end{eqnarray}

Now let us calculate the 4D Planck mass and the 4D effective cosmological
constant. Here we regard the extra dimension up to the first horizon 
at $y_h=c$.  Then, by the integration of the 5D action gives
\begin{eqnarray}
S&=&\int d^4 x\sqrt{-g_4}\int^c_{-c}b(t)dy\,\beta^4\bigg(
\frac{1}{2}\beta^{-2}R_4
-\frac{4}{b^2}\frac{\beta^{\prime\prime}}{\beta}
-\frac{6}{b^2}\bigg(\frac{\beta'}{\beta}\bigg)^2 \nonumber \\
&-&\frac{1}{3}(-\rho+3p+p_5)-\Lambda_b+\frac{2\cdot 4!}{H^2}
-\frac{1}{b}\Lambda_1\delta(y)+{\cal L}_m\bigg)+S_{\rm surface} \nonumber \\
&\equiv &\int d^4 x\sqrt{-g_4}\bigg(\frac{1}{2}M^2_P R_4
-3\Lambda\bigg)
\end{eqnarray}
where ${\cal L}_m=-\rho$ is the Lagrangian for the bulk perfect fluid 
which contributes only at $t=t_0$.
Therefore, the 4D Planck mass is given by
\begin{equation}
M^2_P(t)=\int^c_{-c}b(t)dy \beta^2 
\end{equation}
and the 4D cosmological constant is given by
\begin{eqnarray}
\Lambda(t)&=&\frac{1}{3}\int^c_{-c}b(t)dy\,\beta^4
\bigg[\frac{1}{b^2}\bigg(4\frac{\beta^{\prime\prime}}{\beta}
+6\bigg(\frac{\beta'}{\beta}\bigg)^2\bigg)+\frac{1}{3}(2\rho+3p+p_5) 
\nonumber \\
&+&\Lambda_b+\frac{3\beta^8}{A}+\frac{1}{b}\Lambda_1\delta(y)\bigg].
\end{eqnarray} 
Using the Einstein equations, we can rewrite $\Lambda(t)$ as
\begin{eqnarray}
\Lambda(t)=\frac{1}{3b}\int^c_{-c}dy(\beta^3 \beta')'
+\int^c_{-c}b(t)dy\,\beta^4
\bigg({\bar\Lambda}\beta^{-2}+\frac{1}{9}(2\rho+3p+p_5)\bigg).
\end{eqnarray}  
From the fact that $\beta$ becomes zero at $y_h=c$, 
the resulting $\Lambda(t)$ becomes
\begin{eqnarray}
\Lambda(t)=\int^c_{-c}b(t)dy \bigg({\bar\Lambda}\beta^2
+\frac{1}{9}(2\rho+3p+p_5)\beta^4\bigg).
\end{eqnarray}
Consequently, we get the ratio of the 4D cosmological constant to the 4D
Planck mass which can be interpreted as the time-dependent 
effective curvature, 
\begin{eqnarray}\label{curveff}
\bar\Lambda_{\rm eff}(t)=\frac{\Lambda(t)}{M^2_P(t)}={\bar\Lambda}(t)
+\bigg(\int^c_{-c}dy \beta^2\bigg)^{-1}
\cdot\int^c_{-c}dy\,\frac{1}{9}(2\rho+3p+p_5)\beta^4.
\end{eqnarray}
Therefore, since the bulk matter contributes only at $t=t_0$, 
the difference of the effective c.c for $t>t_0$ from the one for $t<t_0$ 
is given just from ${\bar\Lambda}(t)$ which in our case is 
${\bar\Lambda}_f-{\bar\Lambda}_i$. Note that any value of $\bar\Lambda_f$
is possible.

\section{Vanishing curvature}

In Eq.~(\ref{curveff}), the time-dependence of the curvature is
obtained. It shows the existence of the solution for any instantaneous 
change of the curvature. Therefore, the classical physics does
not determine completely the time dependence of the effective
curvature. It is our hope that some clever action results in a
self-tuning solution with the time-dependence of curvature as
shown in Eq.~(\ref{tdep}).

If classical physics cannot determine the time-dependence, we can
ask a quantum mechanical probability for the transition of the
curvature. Here, we adopt Hawking's Euclidian space
integral for this probability function~\cite{hawking}, from an initial
curvature $\bar\Lambda_i$ to the final curvature $\bar\Lambda_f$.
Since we consider the 5D theory, we must integrate with 
respect to $y$ also up to $y_h$. 
In our notation, the mass dimension of the curvature 
$\bar\Lambda$ is 2, not 4.
Moreover, the Planck mass comes from integration of extra dimension
with the warp factor, so it has a dependence such as
$M_P=M_P(b_f,{\bar\Lambda}_f)$ which is finite for a
vanishing ${\bar\Lambda}_f$ and an infinite $b_f$.
Thus, the probability is estimated to be proportional to
\begin{eqnarray}
\exp \left({\alpha\left[\frac{M^2_P(b_f,{\bar\Lambda}_f)}{{\bar\Lambda}_f}
-\frac{M^2_P(b_i,{\bar\Lambda}_i)}{{\bar\Lambda}_i}\right]}\right)
\end{eqnarray}
where $\alpha$ is a ${\cal O}(1)$ positive numerical number
and $M_P=M_P(b,{\bar\Lambda})$ is finite.
Since this probability function is
infinitely larger for $\Lambda_f=0^+$ compared to any other value of
the final curvature, we obtain by $\sim 100$~\% probability the vanishing
final curvature. In our interpretation
of Hawking's probability, the underlying physics seems to be clear.
We have a definite initial state with $\bar\Lambda_i$ and ask for
the probability of obtaining the final $\bar\Lambda_f$. This
probabilistic determination makes sense only if the classical path
is not determined. If the classical path is determined, the classical
path corresponds to an extremum path.

If we hope to obtain an exit from inflation by a classical
argument only with a self-tuning cosmological constant, 
we need some run-away potential 
for a time-dependent $g_{55}=b(t)$ such that an increasing $b(t)$ gives 
a decreasing cosmological constant. But for a successful interpretation
of cosmology the time dependence must come with a sufficiently large
power $p$ in Eq.~(\ref{tdep}). 

For the strong self-tuning solution, we need a solution allowing
the D-region also to accomodate inflation. 

\section{Conclusion}

We have considered the inflation and the vanishing cosmological
constant in a unified way in 5D self-tuning models. In this framework,
the inflation is designed to originate in the region where only 
the de Sitter space is allowed(D-region). Then, due to the parameter change at
the brane the universe enters into the region where de Sitter, anti-de
Sitter and flat spaces are allowed(F-region). Thus, the initial condition
at the F-region is supposed to be a de Sitter space. 
We considered the time dependence of curvature in the F-region and
proposed that a solution of the cosmological constant problem by equations 
of motion is through the curvature change to zero as $t\rightarrow\infty$.
If the classical equations does not determine the time dependence
of the curvature, a quantum mechanical probability {\it a la} Baum and
Hawking is shown to determine the curvature in the F-region as $0^+$.
This probabilistic solution is clearer in the set-up of this
paper.

\begin{acknowledgments}
JEK is supported in part by the KOSEF ABRL Grant No. R14-2003-012-01001-0,
the BK21 program of Ministry of Education, and Korea
Research Foundation Grant No. KRF-PBRG-2002-070-C00022.
HML is supported by the
European Community's Human Potential Programme under contracts
HPRN-CT-2000-00131 Quantum Spacetime, HPRN-CT-2000-00148 Physics Across the
Present Energy Frontier and HPRN-CT-2000-00152 Supersymmetry and the Early
Universe. HML is also supported by priority grant 1096 of the Deutsche
Forschungsgemeinschaft.
\end{acknowledgments}

\end{document}